\documentclass[11pt]{article}
\usepackage{amsmath, amssymb}
\usepackage{graphicx}
\usepackage{epsfig}
\usepackage{epsf}
\begin{document}

\title{Hartree-Fock Approximation and Entanglement}
\author{Luigi Martina, Giulio  Soliani\\
Dipartimento di Fisica dell'Universita' del Salento { }and Sezione INFN di Lecce.\\
Via Arnesano, CP.193 I-73100 LECCE (Italy)}
\maketitle

\begin{abstract}\textit{The relation between the correlation
energy and the entanglement is analytically  constructed for the
Moshinsky's model of two coupled harmonic oscillators. It turns
out that the two quantities are far to be proportional, even at
very small couplings. A comparison is made also with the 2-point
Ising model. }\end{abstract}

\section{Introduction}
 The concept of entanglement has been recently considered by many
authors [1] in connection with several properties of the quantum
systems and as a potential resource in quantum computation and
information processing. Moreover, entanglement has also been
recognized to play an important role in the study of many particle
quantum systems [2] and experimental measurement have demonstrated
that it can affect macroscopic properties of the condensed matter
[3]. However, a new area of research has been opened by [4], where
it was argued that the entanglement (a quantum observable) can be
used in evaluating the so-called correlation energy: that is the
difference between the true eigenvalue energy of a given molecular
system, composed by identical components, with respect to that one
prescribed by the Hartree - Fock (HF) { }approximation method. In
[5] the case of the formation of the Hydrogen molecule was
discussed, and a qualitative agreement between the von Neumann
entropy of either atom (measuring the entanglement of formation of
the whole system) and the correlation energy as functions of the
inter-nuclear distance was shown. However, the extension of this
idea to multi-atomic molecules and its effectiveness
remains still unclear [6,7]. \\
The correlation energy in this context is an artifact of the
approximation procedure, then it is not a physical observable and,
by second, it can be modified by the adopted method of
calculations. In this sense the conjecture of [5] is not a well
posed question and the answer to it { }must be only qualitative.
Nevertheless, folk says that still entanglement has to play some
role in taking into account the { }difference between the
approximated computation of the (fundamental) energy level for a {
}typical factorized wave function, { }as provided by the
Hartree-Fock approximation, and the {
}$\texttt{"}$true$\texttt{"}$ energy eigenvalue.

This paper is
devoted to a deeper analysis of it, trying to find an operative
and quantitative estimation of the entanglement effects on the
 energy level evaluation by approximated methods. Better, one would like to
 clarify if the correlation energy is in a one-to-one correspondence
 with the entanglement of the fundamental state of bipartite systems.
 To this aim we
analyze a very simple model of coupled harmonic oscillators
introduced by M. Moshinsky about 40 years ago in the paper [8],
and try to arrive to a quantitative measurement of the
entanglement effects on the correlation energy.

In Sec. 2 we briefly review the Moshinsky's model, pointing out
the goodness of the HF approximation for the fundamental state,
both considering the correlation energy and the fidelity with
respect to  the exact solution and for weak couplings of the two
oscillators. In Sec. 3 we evaluate the von Neumann entropy, which
is the unique measure of entanglement for pure states, for the
fundamental state of the Moshinsky's model,  by tracing out the
density matrix operator w.r.t. one of the component oscillator.
Eigenvalues of the reduced density matrix  are computed by
discussing certain integral equations. In order to compare the von
Neumann entropy and the correlation energy, in Sec. 4  we consider
the minimized mean squared deviations of the two quantities. A
discussion of the relative deviations is also performed.
Furthermore an analytic expression of the entropy in terms of the
correlation energy is provided. A comparison with the analogous
expressions for the two point Ising model is discussed. Asymptotic
formulae for very small couplings ( or correlation energy) are
considered at the end. Some final remarks are addressed in the
Conclusions.

\section{The Moshinsky's Model}

In order to evaluate the efficiency and  how good { }the { }HF
mean field method is in computing quantum states, Moshinsky in [8]
proposed a very simple, but non trivial, model of two coupled
spin-\(\frac{1}{2}\) harmonic oscillators in 3 dimensions, in
which all computations can be { }analytically
performed up to the end.\\
In adimensional unities, { }the Hamiltonian of the model reads { }

\begin{equation}\hat{H}=\frac{1}{2}\left(\overset{\rightharpoonup }{p}_1^2+
\overset{\rightharpoonup }{p}_2^2+\overset{\rightharpoonup
}{r}_1^2+\overset{\rightharpoonup
}{r}_2^2\right)+\frac{1}{2}K\left(\overset{\rightharpoonup
}{r}_1-\overset{\rightharpoonup }{r}_2\right)^2,\end{equation}
where \(\overset{\rightharpoonup }{r}_i\) { }and
\(\overset{\rightharpoonup }{p}_i\) denote the position and the
momentum operators of the i-th particle, respectively. { }The
constant \textit{ K} parametrizes the interaction strength of a
supplementary { }harmonic coupling between the two oscillators.
One can interprete such as a model for two { }identical {
}interacting atoms (or molecules). Thus, the coupling would
represent the interatomic interaction, which one can conjecture
weaker than the electron-nucleus interaction. { } Thus, we will
dwell upon the case \textit{ 0 $\leq $ K $\leq $ 1} and the
coupling term in (1) will be considered as a perturbative term. So
changing the value of { }\textit{ K} may be interpreted as a
change into the interatomic interaction. Finally, let us observe
that since we are interested into the fundamental state, the
spinorial aspect of the problem is not relevant.
In fact, the total spin state { }must be singlet. For self-consistency of the present paper, here review the Moshinsky's results. \\
The energy spectrum of the system can be easily computed in the form { } { } { } { } { } { } { } { } { } { } { }

\begin{equation}E_{n, m}=\frac{3}{2}\left(1+(1+2K)^{1/2}\right)+n+m(1+2K)^{1/2},\end{equation}
where\textit{  m} and \textit{ n} are positive integers or 0. For
any \textit{ K { }}the spectrum shows degeneracy, { }partially
broken for { } $ K > 0.$ { }However, the lowest level is always
simple. { }Moreover, eigenvalue crossings occur for higher
eigenvalues at isolated points of \textit{
K . } { } \\
The normalized wave function of the fundamental level is given by { } { } { } { }\textit{  { } { } { } { } { } { } { } { } { } { } { } { } { } {
} { } { } { } { } { } { } { } { } { } { } { } { } { } { } { } { } { } { } { } { } { } { } { } { } { } { } { } { } { } { } { } { } { } { } { } { }
{ } { } { } { } { } { } { } { } { } { } { } { } { } { } { } { } { } { } { } { } { } { } { } { } { } { } }

\begin{equation}\left.\left.\left|\Psi _0\left(\overset{\rightharpoonup }{R},
\overset{\rightharpoonup }{r}\right)\right.\right\rangle =\pi ^{-3/2}
(1+2K)^{3/8}e^{\left.-R^2\right/2}e^{-(1+2K)^{1/2}
\left.r^2\right/2} \left|\frac{1}{2},\frac{1}{2},0,0\right\rangle \right.,
\end{equation} where the mean and relative position   have been defined,
$\overset{\rightharpoonup }{R}\)
=\(\frac{\overset{\rightharpoonup }{r}_1 + \overset{\rightharpoonup
}{r}_2}{\sqrt{2}}$ and
$\overset{\rightharpoonup }{r}\)
=\(\frac{\overset{\rightharpoonup }{r}_1- \overset{\rightharpoonup
}{r}_2}{\sqrt{2}} $ respectively. Looking at the spin state
of { }the system one can see that it is always entangled. So the
question is if also the positional factor has the properties
of an entangled state.\\
Applying the standard HF mean field approximation to  the ground
state, one is led to the wave function

\begin{equation}\left.\left.\left|\Psi _{\text{HF}}\left(\overset{\rightharpoonup }{R},\overset{\rightharpoonup }{r}\right)\right.\right\rangle =
\pi ^{-3/2} (1+K)^{3/4}e^{-(1+K)^{1/2}\left.\left(R^2+
r^2\right)\right/2} \left|\frac{1}{2},\frac{1}{2},0,0\right.\right\rangle
,\end{equation} corresponding to the approximated eigenvalue

\begin{equation}E_{\text{HF}}=3(1+K)^{1/2}.\end{equation}

\begin{figure}
\begin{center}
\epsfig{file= 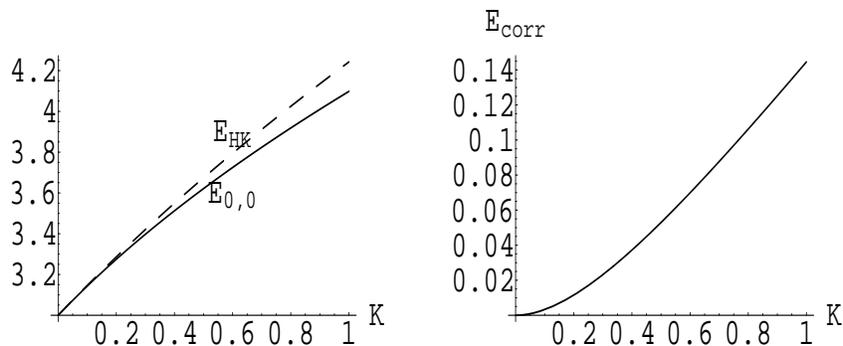, width=12cm,height=5cm}\caption{Exact and
HF approximated fundamental level (on the left).  Correlation
Energy (on the right)}\label{1_gr1}
\end{center}
\end{figure}

\begin{figure}
\begin{center}
\epsfig{file=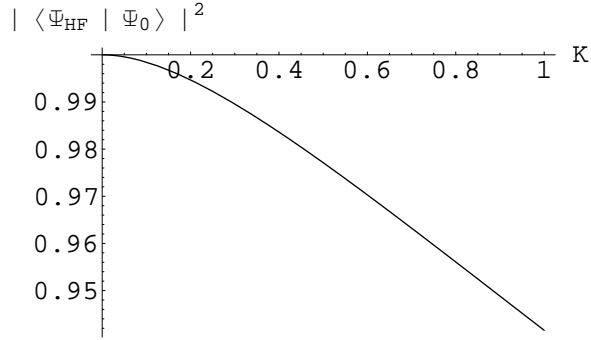, width=8cm,height=5cm}\caption{Overlap of
the HF and exact  ground state.}\label{1_gr2}
\end{center}\end{figure}

Defining the \textit{ correlation energy} as

\begin{equation}E_{\text{corr}}=E_{\text{HF}} - E_{0, 0} = 3 \sqrt{1+K}-\frac{3}{2} \left(1+\sqrt{1+2
K}\right),\end{equation} one obtains a first indication of how
good the HF approximation is. For simplicity this is reported in
Fig. \ref{1_gr1}.
 Of course, by { }Ritz's theorem one has $ E_{\text{HF}} \, \geq
\, E_{0, 0}$, but the approximation looks very good for
small\textit{ K}, indeed they differ at the second order near
\textit{ K }= 0 as
\begin{equation}E_{\text{corr}} =
\frac{3 K^2}{8}-\frac{9 K^3}{16}+O\left(K^4\right),\end{equation}
so that the relative error is less than 3.5$\%$ in the range of $0
\leq K \leq 1$. Moreover, the explicit expression of the overlap (
or the squared fidelity) of the exact and the HF wave function
 $ | \langle \Psi_{\text{HF}}\,|\, \Psi_0 \rangle|^2 $  as a function of $K$ is
\begin{equation}\left|\left\langle
 \Psi _{\text{HF}}|\Psi _0\right\rangle |^2\right.=
\frac{64(1+K)^{3/2}(1+2K)^{3/4}}{\left(K +
\left(1+\sqrt{1+K}\right)\left(1+\sqrt{1+2K}\right)\right)^3},\end{equation}
the graph of which is shown in Fig. \ref{1_gr2}.
It is immediate to see that the distance between the two wave-functions is less than { }95$\%$ for { }\textit{ 0 $\leq $ K $\leq $ 1} .\\
Thus, one can figure out that adding to the HF state further
corrections surely improve the estimations of { }the energy
eigenvalue and increase the fidelity, but the simplest factorized
expression { }in { }(4) will be lost. Indeed, this is the case if
one considers the exact { }expression of { }\(\Psi
_0\)(\(\overset{\rightharpoonup }{R}\),\(\overset{\rightharpoonup
}{r}\)) in (3). { }Qualitatively this means { }that the two
oscillators in the correct fundamental state are entangled,
contrary to what happens in the approximated state. Thus, one would
like to attribute the mismatch between the energy of the two
states { }(i.e. \(E_{\text{corr}}\)) to the entanglement
properties. We will analyze this idea in the next Section.

\section{Entanglement Estimation}

As it is well known,  the main estimator of the entanglement for pure states is the von Neumann entropy { }of the reduced density matrix \(\hat{\rho
}_1\) of one of the component subsystems, i.e.

\begin{equation}S_{\text{vN}} \left[\hat{\rho }_1\right]=-\text{Tr}\left[\hat{\rho }_1 \, \log_{ 2}\hat{\rho }_1\right],\text{  }\text{with}\text{
    }\hat{\rho }_1 = \text{Tr}_2\left[\hat{\rho }\right]\end{equation}
where { }in the second relation one has traced out only with
respect to the second subsystem, when  a proper basis of states
has been chosen in the Hilbert space of the whole system. In the position - spin representation
the complete system density matrix $\rho $ for the fundamental
state (3) is given by

\begin{equation}\left.\hat{\rho _0}=\left|\Psi _0 \right\rangle \left\langle \Psi _0\right|=\left|\frac{1}{2}\right.,\frac{1}{2},0,0\right\rangle
\left\langle \frac{1}{2},\frac{1}{2},0,0\left|\rho
\left(\overset{\rightharpoonup }{r }_1,\overset{\rightharpoonup
}{r}_2,\overset{\rightharpoonup }{r '}_1,\overset{\rightharpoonup
}{r '}_2\right)\right. ,\right.\end{equation}
\begin{equation}\rho \left(\overset{\rightharpoonup }{r }_1,\overset{\rightharpoonup }{r}_2,\overset{\rightharpoonup }{r '}_1,\overset{\rightharpoonup
}{r '}_2\right)= \pi ^{-3}
(1+2K)^{3/4}e^{\left.-\left(R^2+R'^2\right)\right/2}e^{-(1+2K)^{1/2}
\left.\left(r^2+r'^2\right)\right/2},\end{equation} where one has
introduced the supplementary variables \(\overset{\rightharpoonup
}{R'}\) =\(\frac{\overset{\rightharpoonup }{r '}_1+
\overset{\rightharpoonup }{r '}_2}{\sqrt{2}}\) and { }
\(\overset{\rightharpoonup }{r'}\)
=\(\frac{\overset{\rightharpoonup }{r
'}_1-\overset{\rightharpoonup }{r '}_2}{\sqrt{2}}\).

An analogous expression holds for the \( |\Psi
_{\text{HF}}\)$\rangle $ state (4). The main difference between
the two functions is that in the latter case the density matrix
reduces in just the product of gaussian normal distributions in
the real variables (\(\overset{\rightharpoonup }{r
}_1\),\(\overset{\rightharpoonup
}{r}_2\),\(\overset{\rightharpoonup }{r
'}_1\),\(\overset{\rightharpoonup }{r '}_2\)), with the same
variance. Instead, in the case of density matrix (10) the
different coefficients in front of { }(\(R^2\)+\(R'^2\)) and {
}(\(r^2\)+\(r'^2\)) make the presence of mixed (entangling) terms
of the form \(\overset{\rightharpoonup }{r }_1\)$\cdot
$\(\overset{\rightharpoonup }{r}_2\) and { }
\(\overset{\rightharpoonup }{r '}_1\)$\cdot
$\(\overset{\rightharpoonup }{r '}_2\). { }This has consequences
also at level of { }one particle space distribution density, which
is given by

\begin{equation} \rho _{0I }\left( \overset{\rightharpoonup }{r}\right) = \int \rho \left(\overset{\rightharpoonup }{r },\overset{\rightharpoonup }{r}_2,\overset{\rightharpoonup
}{r },\overset{\rightharpoonup }{r }_2\right)
d\overset{\rightharpoonup }{r}_2=\frac{2^{3/2}\text{  }(2
K+1)^{3/4} }{\pi ^{3/2}\left( \sqrt{2
K+1}+1\right)^{3/2}}e^{-\frac{2 \sqrt{2 K+1} r^2}{\sqrt{2
K+1}+1}},\end{equation}
which differs significantly from the one particle density obtained from the HF state, i.e.\\

\begin{equation}\rho _{\text{HF} I}=\frac{ (K+1)^{3/4}}{\pi ^{3/2}}e^{-\sqrt{K+1} r^2},\end{equation}
as, for instance, { }it can be graphically seen in Fig. 3 for a
particular value { }of \textit{ K}.
\begin{figure}\begin{center}
\epsfig{file=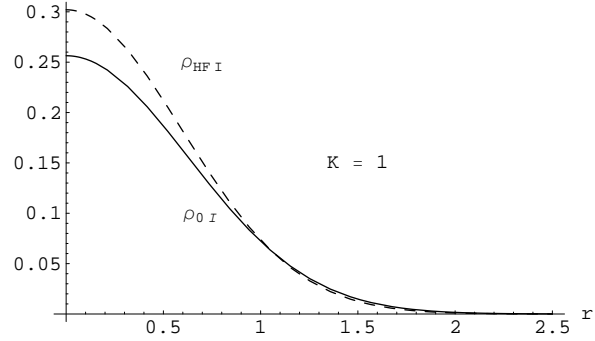,
width=8cm,height=5cm}\caption{Exact and approximated one particle
probability densities, for $K = 1$.}\label{1_gr3}
\end{center}\end{figure}
Thus, in the exact state the mean squared deviation in { }the
position measurements of one particle, regardless the other one,
results { }greater than in the approximated case by a factor
\(\left(\frac{\sqrt{1+K} \left(1+\sqrt{1+2 K}\right)}{2\sqrt{1+2
K}}\right)^{1/2}\). Correspondingly, { }the mean squared
deviations for the modulus of the { }linear momentum { }are
smaller, by the
inverse of the above factor,  in the exact case with respect to the HF one.

The analytic 1-particle spacial density matrix operator
 can be
given in position representation  \(\hat{\rho }_{1 }\) =\( \int
\rho_{1} \left(\vec{r},\, \vec{r'} \right)\; \cdot \; d\,\vec{r'}
\){ }by the kernel
\begin{eqnarray} & \rho _{1 }\left(\overset{\rightharpoonup }{r},\overset{\rightharpoonup }{r'}\right)=\frac{\left(2^{3/2} (2 K+1)^{3/4}\right) }{\left(\sqrt{2
K+1}+1\right)^{3/2} \pi ^{3/2}}\nonumber \\
& \exp \left[\frac{\left(\sqrt{2 K+1}-1\right)^2
\overset{\rightharpoonup }{r}\cdot \overset{\rightharpoonup
}{r'}-\left(K+3 \sqrt{2 K+1}+1\right)
\left(\overset{\rightharpoonup }{r}^{ 2} +\,
\overset{\rightharpoonup }{r'}^{ 2}\right)}{4 \left(\sqrt{2
K+1}+1\right)}\right],\end{eqnarray} which we have to use into
expression (9) in order to compute the corresponding von Neumann
Entropy. { }Actually, one has to take account also of the spin
state. { }But, a well known property of the von Neumann Entropy
says that

\begin{equation}S_{\text{vN}}\left[\hat{ \rho } \otimes  \hat{\sigma } \right] = S_{\text{vN}}\left[\hat{ \rho }\right]\text{  }+\text{  }S_{\text{vN}}\left[\hat{
\sigma }\right],\end{equation} for any factorized { }density
operator { }\(\hat{ \rho }\) $\otimes $ \(\hat{\sigma }\) . {
}Since  this is the case for the system we are considering, where
the reduced matrix in the spin space state is \(\frac{1}{2}\)
\pmb{ 1}  proportional to the identity matrix, the von Neumann
Entropy for the one-particle subsystem can be defined modulo a
constant term { }equal to 1. { }

Furthermore, the expression (9) becomes $ S_{\text{vN}} [\hat{\rho
}_1]= - \sum _{i } \mu _{i }\, {log}_2 \left( { }\mu _{i }\right)$
in terms of the eigenvalues $\mu _{i }$ of the operator $\hat{\rho
}_{1 }$, which is acting on the infinite dimensional space of
squared sommable functions. So, the eigenvalue problem for
\(\hat{\rho }_{1 }\) can be written in the form

\begin{equation}\int \rho _{1}\left(\overset{\rightharpoonup }{r},\overset{\rightharpoonup }{r'}\right)u\left(\overset{\rightharpoonup }{r'}\right)d\overset{\rightharpoonup
}{r'}=\mu \; u\left(\overset{\rightharpoonup
}{r}\right).\end{equation}

As one can see from Eq. (14), the kernel of this { }integral equation is symmetric. Moreover, it is of Hilbert-Schmidt type, since the coefficient
of { }\(\overset{\rightharpoonup }{r'}^{ 2}\) is negative. So the spectrum is real and discrete. Finally, in cartesian coordinates it can be decomposed
{ }in the product of three one-dimensional kernels with the same gaussian structure

\begin{eqnarray} & \tilde{\rho }_ (x, x') = \frac{2^{1/2} (2 K+1)^{1/4}}{\left(\sqrt{2 K+1}+1\right)^{1/2} \pi ^{1/2}}\nonumber \\ &
\exp \left[\frac{\left(\sqrt{2 K+1}-1\right)^2 x\text{
}x'-\left(K+3 \sqrt{2 K+1}+1\right) \left(x^{ 2}+x'^{ 2}\right)}{4
\left(\sqrt{2 K+1}+1\right)}\right]. \end{eqnarray} Accordingly,  the
eigenfunctions of Eq. (16)  can be factorized in the product of
three functions, each of them { }depending only on one real
variable. Thus we put for the eigenfunctions and the eigenvalues

\begin{equation}u_{l, m, n\text{  }}\left(\overset{\rightharpoonup }{r}\right) = w(x) w_{m }(y) w_{n }(z),\text{          }\mu _{l, m, n}= \nu _l
\nu _m \nu _n ,\end{equation} so that the problem is reduced to
solve the integral spectral problem in one dimension

\begin{equation}\int \tilde{\rho }_ (x, x')w_l(x')dx'=\nu _l\text{  }w_l(x) .\end{equation}

Recalling the standard formulae for { }gaussian integrals { }[9]

\begin{equation}\int_{-\infty }^{\infty }x'^i e^{2 q x'- p x'^2} \, dx' =
 \frac{1}{2^{ i-1} p}\sqrt{\frac{\pi }{p}}\text{  }\partial _q^{ i-1}\left(q
e^{\frac{q^2}{p}}\right), \end{equation}
 it is natural to look for solutions of Eq. (19) of the form

\begin{equation}w_l(x) = P_l(x) e^{-\delta  x},\end{equation}
where \(P_l\)(x) is a polynomial of degree \textit{ l} in the variable \textit{ x} { }and $\delta $ is a real positive coefficient.\\
Substitution of the expression (21) in Eq. (19) leads to a pure polynomial expression iff one sets

\begin{equation}\delta = \frac{1}{2} (1+2 K)^{1/4}.\end{equation}

Then, Eq. (19) reduces to \textit{ (l+1)}-dimensional linear eigenvalue problem for the coefficients of \textit{ x} in { }the polynomial { }\(P_l\)(x).
However, the particular form of the coefficients in the r.h.s. of formula (20) makes this problem { }triangular, so one can easily provide the non
degenerate spectrum

\begin{equation}\nu _l = C\text{  }c^{\text{  }l},\end{equation} where
\begin{eqnarray}
 C &=& \frac{2 \sqrt{2} (1+2 K)^{1/4}}{\sqrt{1+K+2 (1+2 K)^{1/4}+3 \sqrt{1+2 K}+2 (1+2 K)^{3/4}}},\nonumber \\
  c &=& \frac{1+K-\sqrt{1+2 K}}{1+K+2 (1+2 K)^{1/4}+3
\sqrt{1+2 K}+2 (1+2 K)^{3/4}}.\nonumber \end{eqnarray}

Of course, these eigenvalues are positive and their sum is equal to 1, because related to a matrix density { }operator. On the other hand, accordingly
to (18) the eigenvalues of the one-particle density matrix (14) are given by

\begin{equation}\mu _k= C^3 c^{\text{  }l + m + n} = C^3 c^k, \; k \,\epsilon\,  \mathbb{N}\cup \{0\},\end{equation}
thus the eigenvalue $\mu _k$ is degenerate of order
\begin{equation}\deg  \left[\mu _k\right]=\frac{k^2}{2}+\frac{3 k}{2}+1.\end{equation}

For comparison in  Fig. 4  we report the graphs of \(\mu _0\)
and { }\(\mu _1\).
\begin{figure}\begin{center}
\epsfig{file=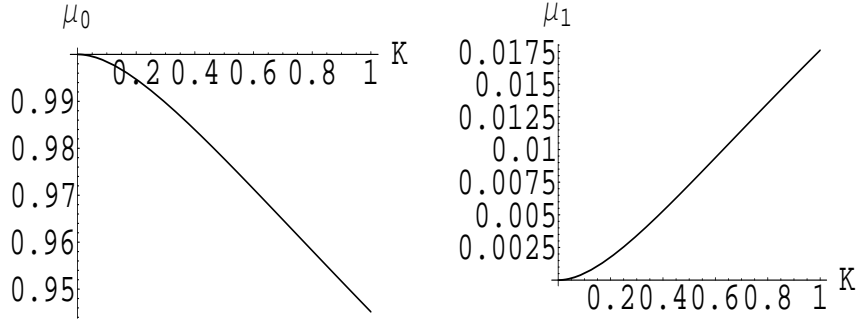,
width=12cm,height=5cm}\caption{The eigenvalues $\mu_0$ and $\mu_1$ of the reduced density operator ${\hat \rho}_1$.}\label{1_gr4}
\end{center}\end{figure}
Hence, if we are allowed to interprete the eigenvalues of the
density matrix operator $\hat \rho_1$
 as { }the probabilities to find the one particle subsystem in the states of { }a \textit{
K}-parametrized family of harmonic oscillators, for the
fundamental one it is very close to 1 and  slowly decreasing in
\textit{ K}. { }At K $\approx $ 1, the probability to find it in
the first excited state is about 1.76$\%$ and exponentially
smaller for the highest excited states. { }The lack of coherence
can be estimated also by computing the Trace $[ \hat{\rho }_1^2
]$. Indeed this quantity is $1$ only if it corresponds to a pure
state, but in the present case one has
\begin{equation}\text{Tr}\left[\hat{\rho }_1 ^2\right]= \frac{8 (1+2 K)^{3/4}}{\left(1+\sqrt{1+2 K}\right)^3}, \end{equation}
which is a decreasing function in \textit{ K}, as can be seen from
Fig. 5.
\begin{figure}\begin{center}
\epsfig{file=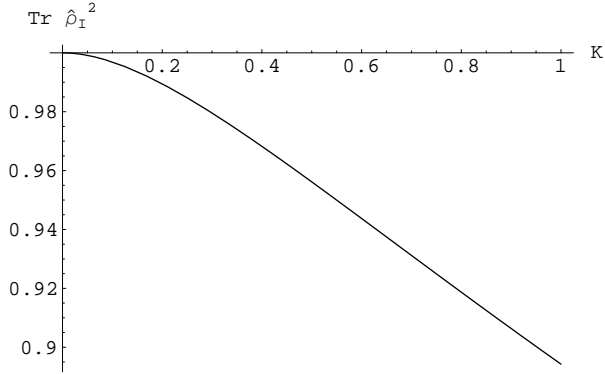, width=8cm,height=5cm}\caption{The trace of
${\hat \rho}_1^2$, as a function of the coupling constat $K$. The
state of one the Moshinsky's oscillators becomes more mixed for
increasing $K$. }\label{1_gr5}
\end{center}\end{figure}

Now, we are in the condition to compute the von Neumann entropy,
by rewriting formula (9) as
\begin{eqnarray} & S_{\text{vN}} \left[\hat{\rho }_{1}\right] = - \sum _{k=0}^{\infty } \deg \left[\mu_k \right] \mu_k \log_2 \mu _k =\nonumber
\\ \\
& \frac{3}{ln[4]} \left\{2 \left( 1 + \chi^2 \right) \ln\left[ 1 +
\chi \right] - 2 \chi \,ln\left[ 4 \chi \right]-\left(-1+\chi
\right)^2 \ln\left[-1 + \chi^2
\right]\right\}\chi^{-1},\nonumber\end{eqnarray} where
\begin{equation}\text{      $\chi $ = }(1+2 K)^{\frac{1}{4}}\text{. }\end{equation}
A graphic of $ S_{\text{vN}} [{\hat {\rho }}_{1}]$
\begin{figure}\begin{center}
\epsfig{file=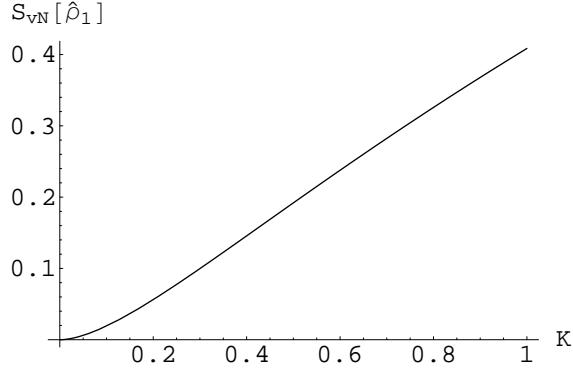, width=8cm,height=5cm}\caption{The
entanglement of the Moshinsky's model as a function of the
coupling constat $K$ }\label{1_gr5bis}
\end{center}\end{figure}
is given in Fig. 6 for the range of $ 0 \leq  K  \leq 1$.

Although in the present case the entropy is not upper bounded,
since we are dealing with a system with an infinite-dimensional
Hilbert space, the expression (27) is very to
 the analogous one for the 2-point Ising model
[5,10], for which { }the 1-particle von Neumann entropy and the
correlation energy are
\begin{eqnarray} & S_{\text {vN}}^{2-\text{Ising}} \left[\hat{\rho }_{1}\right] = - \frac{1}{\sqrt{4+\lambda ^2}
\,ln[4]} \label{IsingEntropy}\\ & \left\{\left(-2+\sqrt{4+\lambda
^2}\right) ln\left[\frac{1}{2}-\frac{1}{\sqrt{4+\lambda
^2}}\right]+\left(2+\sqrt{4+\lambda ^2}\right)
ln\left[\frac{1}{2}+\frac{1}{\sqrt{4+\lambda
^2}}\right]\right\},\nonumber\\ & E_{\text{corr}}=\sqrt{4+\lambda
^2}-2\text{ }\label{IsingEn},\end{eqnarray} where $\lambda $ is
the normalized ferromagnetic coupling.

 For large $
K$   the entropy $ S_{\text{vN}} \left[\hat{\rho }_{1}\right]$
increases logarithmically according to the expansion at
infinity $$ S_{\text{vN}} \left[\hat{\rho }_{1}\right]\approx
\frac{3 \, ln [K]}{4 \, ln [2]} - \frac{21}{4} + \frac{3}{ln (2)
}+ O\left[\frac{1}{ K^{1/2}}\right].$$

On the other hand, the  behavior of $S_{\text{vN}} \left[\hat{\rho
}_{1}\right]$ near K = 0 can be described by its series expansion
\begin{eqnarray} S_{\text{vN}} \left[\hat{\rho }_{1}\right] =  \left(\frac{3}{4} + \frac{3}{16
   \, ln (2)}\right) K^2 -\frac{3 K^3}{2} + \left(\frac{177}{64}-\frac{189}{512 \, ln (2)}\right)
   K^4  \nonumber \\ - \frac{3 K^2 \left(16 - 32 K + 59 K^2\right)
   }{128 \, ln (2)} \, ln (K) + O\left(K^5\right),\end{eqnarray}
but this approximation becomes inaccurate very rapidly. From the
above expression one sees the asymptotic behavior of the entropy
close to 0 is controlled by a logarithmic term, differently from
the correlation energy (6), which has a pure power expansion.
Then, we cannot expect a great similarity between the two
functions, also at very small values of K.

\section{Comparing Entropy and Correlation Energy}

The comparison of entropy and correlation energy requires at least
a common scale of unities, the former being a pure number. So, the
simplest proposal is { }to study a family of deviation functions
of the form
\begin{equation}\Delta (\alpha ,K)=\alpha  S_{\text{vN}} \left[\hat{\rho }_{1}\right]- E_{\text{corr}} ,\end{equation}
where $\alpha $ is a parameter to be properly chosen. The choice
of $\alpha $ can be done in several ways but, trying to minimize
the values of { } $\Delta $($\alpha $) in a range of \textit{ K,}
a criterion could be given by the minimization of the squared
deviations functional
\begin{equation}I[\alpha ]=\int_0^1 \Delta (\alpha ,K)^{ 2} \, dK. \end{equation}
Quite arbitrarily, the domain of integration  is determined  by
our interest to small perturbations of the 2-independent harmonic
oscillators system. This is equivalent to set
\begin{equation}\alpha _{\min } = \frac{\int _0^1 S_{\text{vN}} \left[\hat{\rho }_{1}\right]
E_{\text{corr}}dK}{\int _0^1 S_{\text{vN}} \left[\hat{\rho
}_{1}\right]^2dK} \approx \text{0.318949},\end{equation} where a
numerical evaluations is needed because of the intricate form of
the integrand functions. The graphic of the function $\Delta
$(\(\alpha _{\min }\),K) is given in Fig. 7 .
\begin{figure}\begin{center}
\epsfig{file=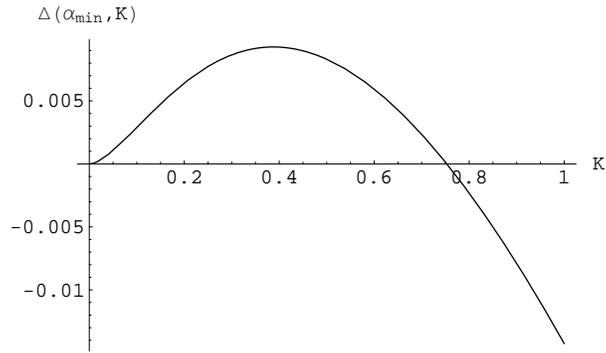, width=8cm,height=5cm}\caption{A plot of the
minimized  deviations  between entropy and correlation energy.
}\label{1_gr6}
\end{center}\end{figure}
 An estimation of the relative deviations of the correlation energy with respect to the entropy is given in Fig. 7. In both cases we have considered
the relative deviation at the minimum of the squared deviation
functional (31).
\begin{figure}\begin{center}
\epsfig{file=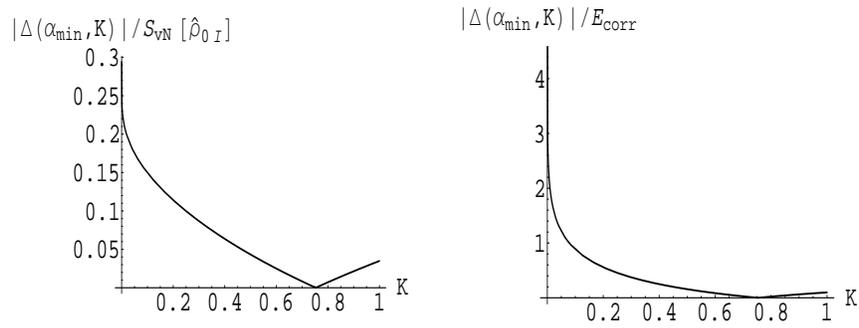, width=12cm,height=5cm}\caption{The relative
minimized  deviations  between entropy and correlation
energy.}\label{1_gr7}
\end{center}\end{figure}
As one can see, in both cases the relative error is quite big, {
}particularly { }close to 0, where the behavior of these functions
are largely controlled by  logarithmic or cuspidal singularities.
In particular, while in the expansion of $\Delta $(\(\alpha _{\min
}\),K)/ \(S_{\text{vN}}\) [\(\hat{\rho }_{1}\)]  a finite cuspidal
singularity appears at 0, i.e.
\begin{equation}\frac{\left|\Delta \left(\alpha _{\min },K\right)\right|}{S_{\text{vN}} \left[\hat{\rho }_{1}\right]}
= \text{0.035439}+O(K),\nonumber \quad \frac{d}{d K} \left(
\frac{\left|\Delta \left(\alpha _{\min
},K\right)\right|}{S_{\text{vN}} \left[\hat{\rho }_{1}\right]}
\right)  \simeq  \frac{-8.20851 \times 10^{-16}}{K^4},
\end{equation}
 while the relative deviation with respect to the correlation energy shows
  a true logarithmic divergence as
\begin{equation}\frac{\left|\Delta \left(\alpha _{\min },K\right)\right|}{ E_{\text{corr}}} =
1+1.14945\times 10^{-16} \ln \left(K\right) +O(K),\end{equation}
making very questionable the meaning of these quantity, taking
into account also the tiny absolute values of the expansion
coefficients.

 On the other hand, { }correlation energy and entropy
look almost proportional in the restricted range { }$\sim $
0.6$\leq $ \textit{ K} $\leq $ 0.8. In this region of the \textit{
K} parameter it is confirmed in explicit way how the correlation
energy is a direct effect of the entanglement. In general such a
relation is less obvious. In fact, first let us observe that both
functions \(S_{\text{vN}}\) [\(\hat{\rho }_{1}\)]  and
\(E_{\text{corr}}\) are both  monotonically increasing in \textit{
K}. { }Therefore, since \(E_{\text{corr}}\) is a quite simple
algebraic function on the coupling constant, it can be easily
inverted and replaced into \(S_{\text{vN}}\) [\(\hat{\rho }_{1}\)]
, obtaining again a one-to-one correspondence, say {
}\(\tilde{S}_{\text{vN}}\) (\(E_{\text{corr}}\)), the expression
of which is a little involved,  actually
\begin{eqnarray} & \tilde{S}_{\text{vN}}  \left(E_{\text{corr}}\right)  =  \frac{1}{\sqrt{2
   \tau +3} \, ln (4)}  \left\{4 \sqrt{3} (\tau
   +3) \, ln \left(\sqrt{6 \tau +9}+3\right) \right. \\ & -2 \sqrt{3} \, ln (6) \, (\tau +3)-2 \left(\sqrt{3} (\tau
   +3)-3 \sqrt{2 \, \tau + 3}\right) \, ln (\tau ) \nonumber\\ &\left.
   -3 \sqrt{2 \, \tau
   +3} \; ln (2\, \tau +3)-3 \sqrt{2 \tau +3} \, ln
   (12)\right\},\nonumber
\end{eqnarray}
where we have introduced the auxiliary variable $\tau =
\textit{E}_{\textit{corr}}+\sqrt{2}
\sqrt{\textit{E}_{\textit{corr}}
   \left(\textit{E}_{\textit{corr}}+3\right)}$. A plot of this function is given in Fig. 9.
\begin{figure}\begin{center}
\epsfig{file=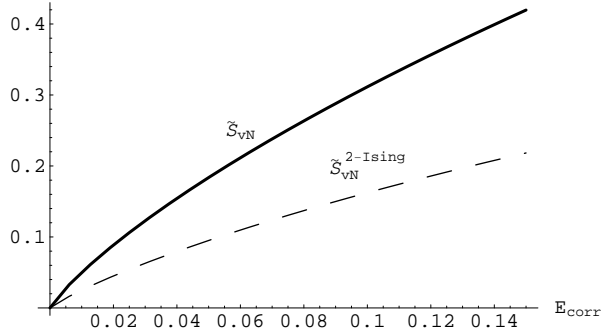, width=8cm,height=5cm}\caption{The
entanglement as a function of the correlation energy for the
Moshinsky's model (continuous  curve) and for the 2-point Ising
model (dashed curve).}\label{1_gr7}
\end{center}\end{figure}
The analogous expression for the 2-point Ising model  results
directly by substituting $E_{corr}$  given in Eq. (30) into the
expression (29) for the entanglement.  A  plot of this function is
as is also sketched in Fig. 9.

In both cases the entanglement is an increasing function of the
correlation energy. In particular, one can search  explicit but
approximated expressions of the entanglement for small values of
$E_{corr}$, corresponding to small couplings. Indeed,   {
}including logarithmic corrections at the lowest order near
\(E_{\text{corr}}\) $\approx $ 0, { }one obtains { }
\begin{eqnarray} \tilde{S}_{\text{vN}}  \left(E_{\text{corr}}\right)\approx
 \frac{ \left(1+\,ln[6]-ln\left[E_{\text{corr}}\right]\right)}{2
\,ln[2]} E_{\text{corr}} ,\\ \tilde{S}_{\text{vN}}^{
2-\text{Ising}} \left(E_{\text{corr}}\right) \approx \frac{
\left(1+2 \,ln[2]-ln\left[E_{\text{corr}}\right]\right)}{4
\,ln[2]} E_{\text{corr}} ,\end{eqnarray} for the Moshinsky's
oscillators and the 2-point Ising model,  respectively. As one can
see, the two asymptotic formulae above are very similar, but not
sufficient to argue that there exists a general argument allowing to
compute the coefficients appearing in the above developments.

 For large perturbations, or equivalently for $
E_{\text{corr}} \rightarrow \infty $, one has the asymptotic
expansions
\begin{equation} \tilde{S}_{\text{vN}}
\left(E_{\text{corr}}\right) =  \frac{3 \left(2+\sinh ^{-1}(1)-\,
ln (24)\right)}{\, ln (4)} + \frac{3 \, ln
\left(E_{\text{corr}}\right)}{2 \,
ln(2)}+O\left(\frac{1}{E_{\text{corr}}}\right)\end{equation} for
the Moshinsky's model, while for the finite levels 2-point Ising
model the entanglement will approach the limiting value 1 as
\begin{equation} \tilde{S}_{\text{vN}}^{ 2-\text{Ising}} = 1-\frac{2}{\, ln (2)
   E_{\text{corr}}^2}+O\left(\frac{1}{\text{E}_{corr}^3}\right).  \end{equation}

Concerning  the singular behavior of the entanglement as a
function of the correlation energy it does not seem related to the
specific way of its estimation. In fact, in alternative to the von
Neumann entropy one can use the concurrence [10], which  uniquely
defines the entropy at least for the Ising model (see Eq.
\ref{IsingEntropy}). In  this case
 the concurrence takes the form
  \begin{equation} C^{2-Ising} = \frac{\lambda }{\sqrt{\lambda ^2+4}} \label{concurrIsing}.
  \end{equation}  Furthermore,
  expressed in terms of the correlation energy  in (\ref{IsingEn}), it becomes
 \begin{equation} C^{2-Ising} = \frac{\sqrt{E_{\text{corr}}\left(E_{\text{corr}}+4\right)}}{E_{
 \text{corr}}+2}.\label{concurrIsingEn} \end{equation}As one can see, the first derivative of
 this expression contains an algebraic singularity at 0 instead of a logarithmic one,
   as it shown by Eq. (\ref{concurrIsingEn}), and it has a  monotonic algebraic increasing to the limiting value 1.

Inspired by [10], one can define  the analogous of the concurrence
for the Moshinsky's system  (at least limiting ourselves to the
set of considered ground states) as
\begin{equation} \emph{C} = 1 - Tr[{\hat \rho}_1^2], \label{ConcMo} \end{equation}
which takes values in the range $\left[ 0, 1 \right[ $ . It  is
invariant under local unitary transformations  on the separate
oscillators (reduced to changes of phases) and provides a unique
mapping for \(S_{\text{vN}}\) [\(\hat{\rho }_{1}\)], by using Eq.
(28) into (26) and, then, replacing into (27).

As a function of the correlation energy, the above defined
quantity takes the form
\begin{equation} \emph{C}\left( E_{corr} \right) = 1- \frac{3 \sqrt{3} \left( 2 E_{corr} + 2 \sqrt{2} \sqrt{E_{corr}
   \left(E_{corr} +3 \right)} + 3\right)^{3/2}}{\left(E_{corr}+\sqrt{2}
   \sqrt{E_{corr} \left(E_{corr}+ 3\right)}+3\right)^3},
\label{ConcMOEn}\end{equation} which is regular in the origin, but
it is not its second derivative. Again a singularity is signaling
the a greater rate of increase of the entanglement  with respect
to the correlation energy, for small values of the latter.

One advantage of the expression (\ref{ConcMOEn}) is that it can be
easily inverted, providing a quite simple expression of the
correlation energy of the Moshinsky's model in terms of the
entanglement, which was one of the original motivations of the
present work. Specifically, one has the correlation energy as a
function of the concurrence (41)
\begin{eqnarray} E_{corr} &= \frac{3}{(1-\text{C})^{5/6}} \left[\sqrt{2 \text{C}-\sqrt[3]{1-\text{C}} \left( 2 \sqrt{(1-\text{C})^{4/3}-3
   (1-\text{C})^{2/3}+2} - 3 \right)
  - 2}  \right. \nonumber\\ &\left. + \, (1-\text{C})^{1/6} \left( \sqrt{(1-\text{C})^{4/3}-3
   (1-\text{C})^{2/3}+2}- 1\right) \right] . \label{EnConcMo}\end{eqnarray}
In particular, the power expansion of the above expression near 0
concurrence is given by
\begin{equation} E_{corr} \approx \text{C}+\sqrt{\frac{2}{3}} \,
\text{C}^{3/2}+\frac{2\,
   \text{C}^2}{3}+O\left(\text{C}^{5/2}\right).\label{EnConcMoApp}\end{equation}
This expression gives a direct relation between the correlation
energy and  the entanglement via the quantity defined in
(\ref{ConcMo}), much simpler than by using the von Neumann
entropy.

\begin{figure}\begin{center}
\epsfig{file=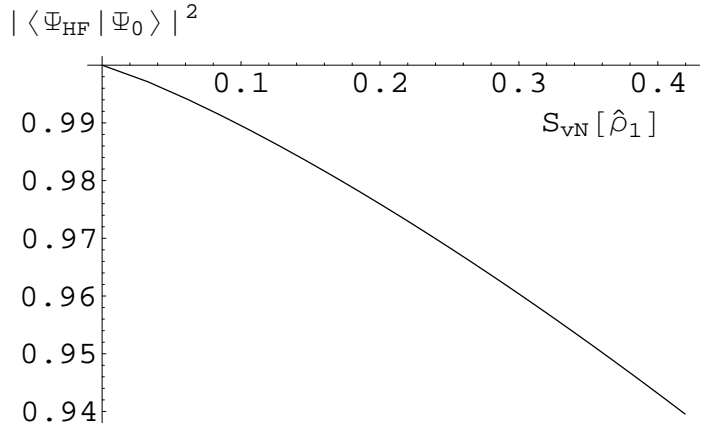, width=10cm,height=6cm}\caption{The overlap
for HF w.r.t. the exact
 ground state of the Moshinsky's model
entanglement as a function of the }\label{SF}
\end{center}\end{figure}

Finally, the fidelity of the fundamental state of the Moshinsky's
model with the corresponding HF state, or equivalently the overlap
(8),  can expressed as a function of  the entanglement. In some
sense, we are comparing two different way to measure the
"distance" between the two curves of states, even if neither
quantities actually have the properties of a distance. However,
also in this case a monotonic function can be obtained for any
pair of states corresponding to the same coupling constant $K$, or
correlation energy $E_{corr}$. But, in general, an explicit
expression of such a function is unknown for generic states of the
Moshinsky's model.  A plot of the function  is shown in Fig.
\ref{SF}.
Also for this particular situation, even if regular, the behavior
at the origin  possesses a singularity in its higher derivatives.
\section{Conclusions}

In the present paper, from point of view of the entanglement
theory of quantum systems,  we have described the relations
occurring between two special curves in the space states of a
family of bipartite system (the Moshinsky's model of harmonic
oscillators),  continuously parametrized by a coupling constant.
To varying of that, one curve contains the exact ground states,
the other the states provided by the HF method approximated
solutions. Of course, for vanishing coupling the two curves flow
out from the same state, but we would characterize how they
separate in terms of entanglement.  The peculiarity of the latter
curve is to lie always in the manifold of 0 entanglement, while
the other one suffers a monotonic increasing entanglement. This
could provide a measure of the "distance" between the two curves
and, maybe, of the geometry of the space state around them. On the
other hand, a similar, but not necessarily equivalent description
is given in terms of the correlation energy, which in principle is
defined only for pairs of corresponding states (at same coupling
$K$) in the two curves. We have proved that entanglement and
correlation energy are one-to-one along these curves. However some
peculiarities arise. First, they are far to be proportional and
only for certain intervals of the coupling constant their squared
deviations can be considered small to some percentage. Second, at
origin they have a quite different rate of increasing. We show
that this phenomenon occurs not only in terms of the von Neumann
entropy, but also by introducing an adaptation of the concurrence.
From the view point of physico-chemical calculations the above
observations say that an artifact of the calculation methods, the
correlation energy, can be interpreted and calculated in terms of
the entanglement of the wave function of the exact solution.
Moreover, we have quite simple algebraic relations in terms of the
concurrence (see Eq. (\ref{EnConcMoApp})). However, at the moment
we have not a general method to compute directly the coefficients
of such type of expansions. These could be very useful in order to
have an alternative a priori estimation of the error made in
computing the correct expectation values of the energy.

We have shown  also that the overlap of the exact wave functions
with the corresponding  HF approximations are in a one-to-one
correspondence with the correlation energy and the entanglement.
In particular a monotonic decreasing of the overlap occurs as a
function of the entanglement, with a non analytic behavior at the
origin. Differently as above, such a type of relation may be
useful in the estimation of the entanglement, which could be quite
complicated to compute for bipartite systems with high inner
degrees of freedom, as the dimers of complex molecules (see [6-7]
for instance).

Finally, we would stress that if a general principle of nature
would state  that it is impossible to create (or increase)
entanglement between remote quantum systems by local operations
[11] (in analogy with the second law of the thermodynamics),
expressions like (32) may suggest proper modifications in presence
of globally controlled operations, analogous to the isothermal
transformations. In this respect, the true meaning of the
minimizing parameter $\alpha$ is still obscure and could be a
further direction of investigation. It could be just a case but,
as noticed in [7], the properties of the function $ \Delta
\left(\alpha _{\min },K\right)$ could be used to predict special
configurations of the considered physical systems, in a sort of
balance between entanglement and energy.

\subsubsection*{Acknowledgements}
The authors acknowledge the Italian Ministry of Scientific
Researches (MIUR) for  partial support of the present work under
the project SINTESI 2004/06 and the INFN for  partial support
under the project Iniziativa Specifica LE41.

\subsection*{References}

\begin{itemize}
\item[[1]] M. A. Nielsen and I. L. Chuang : " Quantum Computation
and Quantum Information ", Cambridge University Press, Cambridge
(2000).\\
\item[[2]] Y. Chen, P. Zanardi, Z.D. Wang, F.C. Zhang: "
Entanglement and Quantum Phase Transition in Low Dimensional Spin
System ", quant-ph/0407228.\\
\item[[3]] L. He, G. Bester and
A. Zunger, cond-math/0503492 (2005).\\
\item[[4]] D. M. Collin, {\it Z. Naturforsch.}  {\bf A48}, 68 (1993). \\
\item[[5]] Z. Huang, S. Kais:'' Entanglement as measure of the
electron-electron correlation in quantum chemistry calculations'',
\textit{ Chem.Phys.Lett.}
\pmb{413}, 1 (2005).\\
\item[[6]] T. Maiolo, F. Della Sala, L. Martina, G. Soliani, quant/ph/0610238, to appera in {\it Theor. Math. Phys.} (2007)\\
\item[[7]] T. Maiolo, L. Martina, G. Soliani,
quant/ph/0704.0520v1.  [quant-ph]
\item[[8]] M. Moshinsky:$\texttt{"}$ How Good is the Hartree-Fock Approximation ?$\texttt{"}$, { }\textit{ Am. J. Phys}. \pmb{ 36},  52 (1968).\\
\item[[9]] { }I. S. Gradshteyn and I.M. Ryzhik : $\texttt{"}$
Table of integrals, series, and products$\texttt{"}$ ; Alan
Jeffrey, editor , { }Academic Press, San Diego (1994).\\
\item[[10]] S. Hill and W. K. Wootters, {\it Phys. Rev. Lett.},
{\bf 78}, 5022 (1997);  W. K. Wootters,  {\it Phys. Rev. Lett.},
{\bf 80}, 2245 (1998).
 \item[[11]] S. Popescu, D. Rohrlich, {\it Phys. Rev.} {\bf A 56},
 R3319 (1997).
\end{itemize}

\end{document}